\documentclass[aps,twocolumn,groupedaddress,showpacs]{revtex4}

\usepackage{graphicx}
\usepackage{epsfig}

\begin{document}
\bibliographystyle{apsrev}

%\preprint{}

\title{Oscillatory nonlinear conductance of an
interacting quantum wire with an impurity}

\author{Fabrizio~Dolcini$^1$, Hermann~Grabert$^1$, In\`es~Safi$^2$, and
  Bj\"orn~Trauzettel$^{1,2}$}

\affiliation{${}^1$ Physikalisches Institut,
Albert-Ludwigs-Universit\"at,
 79104 Freiburg, Germany\\
${}^2$Laboratoire de Physique des Solides, Universit\'e Paris-Sud,
91405 Orsay, France}

\date{August 8, 2003}

\begin{abstract}

The nonlinear conductance  of a one-dimensional quantum wire
adiabatically coupled to Fermi liquid electron reservoirs is
determined in presence of an impurity. We show that
electron-electron interaction in connection with the finite length
of the wire leads to characteristic oscillations in the current as
a function of the applied voltage.

\end{abstract}

\maketitle

It is well known that Fermi Liquid (FL) theory is not appropriate to
describe transport properties of one-dimensional (1D) electron
systems. Correlation effects completely destroy Landau's
quasiparticle picture, and the elementary excitations of a 1D
quantumwire (QW) can be modelled in terms of bosonic plasmon and
spin density modes. An equivalent representation employs
fractionally charged quasiparticles \cite{fisherGlazman,lederer}.
The charge $ge$ depends on the parameter $g < 1$ characterizing
the electron-electron interaction strength, where $g=1$
corresponds to the non-interacting limit and $g \to 0$ means
strong correlations. An impurity potential in a 1D QW is strongly
renormalized by the interaction \cite{kane92} and effectively acts
as a back-scatterer of these fractionally charged
quasiparticles~\cite{fisherGlazman}.

These well known properties of 1D fermions are a natural
consequence of the Luttinger Liquid (LL) model
\cite{haldane,gogol98}, which generically describes the low energy
physics of QWs. Despite these drastic departures from FL
theory, the strong correlation effects are often disguised in
transport measurements. For instance, the conductance of a clean
QW is not  affected at all by the electron-electron interaction
within the wire \cite{safi95,maslo95}, and the shot noise of a
wire with an impurity does not reflect the fractional charge of
the backscattered quasiparticles \cite{ponom99,trauz02}. The
reason lies in the coupling of the QW to the FL leads at
its ends: experimentally only the chemical potential of electrons
in the leads can be manipulated, while the chemical potential of
the fractionally charged quasiparticles is not controllable. This
represents the main difference with respect to the case of edge
currents in fractional quantum Hall bars, where right and left
movers are spatially separated (chiral LL), and the chemical
potential of the Laughlin quasiparticles can be controlled by the
upstream reservoir \cite{milli96}. In contrast, in a semiconductor
QW or a nanotube right and left movers interact and interfere in
the same channel resulting in a nontrivial influence of the
contacts even for the case of ideal adiabatic coupling.

Due to the important effects of the leads, a finite length QW in
many respects acts as a complicated scatterer in a FL wire, and
transport properties obey FL behavior. Thus, experimental
observations of LL behavior in semiconductor QWs
\cite{taruc95,yacob96} and nanotubes \cite{dekker} are rather
based on the temperature dependence of the conductance, or on
modifications by the Coulomb interaction of resonant tunnelling
through quantum dot structures \cite{doublekink}. Here we show
that due to the interaction within the wire, the {\em nonlinear}
conductance of a QW with a single impurity shows oscillatory
behavior. In contrast to oscillations in double impurity systems
\cite{doublekink} or in wires with tunneling contacts to the leads
\cite{liang01,peca03}, the oscillations predicted here are
completely absent in a non-interacting wire. So we do not discuss
modifications of an oscillatory effect already present in FL
theory, but non-monotonic behavior of the conductance that is a
signature of LL behavior.

To study a QW of finite length coupled to non-interacting
reservoirs, we employ the inhomogeneous LL model
\cite{safi95,maslo95}, where the interaction parameter has a
spatial dependence. It varies from $g(|x|>L/2)=1$ in the leads to
$g(|x|<L/2)=g<1$ in the QW. Although in real physical systems
without backscattering at the contacts the variation of $g(x)$
from unity to $g$ must be smooth on a length scale $L_s \gg
\lambda_F$, where $\lambda_F$ is the Fermi wavelength, it turns
out that, as long as $L_s \ll L$, the simpler model of a $g(x)$
step-function already captures the important physical properties
of a QW adiabatically connected to non-interacting electron
reservoirs. Moreover, we have to keep in mind that only under the
assumption $L_s \ll L$ the QW has a well-defined length $L$.
Throughout this article we treat the spin-polarized situation at
zero temperature, leaving finite temperature aspects for a future
publication.

For $\hbar \equiv 1$, the full Lagrangian of our model is
$\mathcal{L}=\mathcal{L}_0+\mathcal{L}_I+\mathcal{L}_U$. In
bosonization, see Ref.~\cite{gogol98} for a recent review, the
inhomogeneous LL Lagrangian of a clean system in terms of the
phase field $\theta(x,t)$  reads
\begin{equation} \label{h_0}
\mathcal{L}_0 = \frac{1}{2} \int dx \left[ \frac{1}{v_F}
\left(\partial_t \theta   \right)^{2} - \frac{v_F}{g^{2}(x)}
\left(\partial_x \theta   \right)^{2} \right] \; ,
\end{equation}
where $v_F$ is the bare Fermi velocity of non-interacting right-
and left-movers. The backscattering of an impurity at position $x_0$ yields
the term
\begin{equation} \label{h_i}
\mathcal{L}_I=-\lambda \cos\left[\sqrt{4\pi} \theta(x_0,t) +2 k_F
x_0 \right] \label{LI}
\end{equation}
with the `bare' impurity strength $\lambda$, whereas the impurity
forward scattering omitted here does not affect the
current-voltage characteristics. A third term
\begin{equation} \label{h_u}
\mathcal{L}_U = -\frac{e}{\sqrt{\pi}} \int dx \, U(x) \partial_x
\theta \; ,
\end{equation}
describes the coupling to an electrostatic potential $U(x)$, with
$eU(x<-L/2) = \mu_L$, $eU(x>L/2) = \mu_R$, and $U(x)=0$ for
$|x|<L/2$. Here, $\mu_L$ and $\mu_R$ are respectively the chemical
potentials in the left and right reservoir, and we assume a
symmetrically biased system with $\mu_L = eU/2$ and $\mu_R =
-eU/2$. $U(x)$ describes the shift of the band bottom due to
electroneutrality in the FL leads \cite{trauz02}. In
contrast, the QW itself does not remain electroneutral in presence
of an applied voltage $U$ and the electrostatics within an
interacting QW emerges naturally as a steady-state effect
\cite{egger96,trauz02}.\\

In bosonization, the average current is given by $\langle I
\rangle = - (e/\sqrt{\pi}) \partial_t \langle \theta(x,t)
\rangle$, and it is most convenient to decompose the field
operator $\theta(x,t)$ as \cite{egger96}
$\theta(x,t)=\theta_p(x,t)+\phi(x,t)$, where $\theta_p(x,t)$ is a
particular solution of the equation of motion following from the
action
$\mathcal{S}[\theta]=\mathcal{S}_0+\mathcal{S}_I+\mathcal{S}_U$
that corresponds to Eqs.~(\ref{h_0}) to (\ref{h_u}). On the other
hand, $\phi(x,t)$ can be regarded as the fluctuation operator
which therefore has to fulfill
\begin{equation} \label{relation}
\partial_t \langle \phi(x,t) \rangle = 0 \; .
\end{equation}
Straightforward algebra shows that the particular solution may be
written as
\begin{equation}
\theta_p(x,t) = - \frac{e (U-V)}{2\sqrt{\pi}} t +  \left\{
\begin{array}{r@{\quad \;, \quad}l} \frac{e(U-V)}{2\sqrt{\pi}v_F}|x| & |x|>
  \frac{L}{2} \\
-\frac{eg^2V}{2\sqrt{\pi}v_F}|x| & |x|< \frac{L}{2}
\end{array} \right. .
\end{equation}
Consequently, the average current reads
\begin{equation}
\langle I \rangle = \frac{e^{2}}{2\pi} (U-V) \; , \label{I}
\end{equation}
where $V$ is a parameter corresponding to the four-terminal
voltage drop at the impurity site. In order to determine $V$, one
has to impose the condition (\ref{relation}) defining the
fluctuation field $\phi$.

The effective action for $\phi$ is easily found to read
\begin{eqnarray}
S[\phi] &=& \frac{1}{2} \int dt \int dx \left[ \frac{1}{v_F}\left(
\partial_t \phi \right) ^2 - \frac{v_F}{g^{2}(x)}
\left(\partial_x \phi   \right)^{2} \right] \nonumber \\
&-& \frac{eV}{\sqrt{\pi}} \int \, dt \phi(x_0,t) \label{Seff} \\
&-& \lambda \int dt \cos \left[\sqrt{4\pi}\phi(x_0,t) - e(U-V)t+2
k_F x_0 \right]  \nonumber  .
\end{eqnarray}
For weak backscattering the terms related to the impurity, namely
the second and third line of Eq.~(\ref{Seff}), can be treated as
perturbations on the free boson part (first line) employing
standard techniques. The conditions when such a perturbative
approach is reliable will be discussed below. It  is convenient to
characterize the conductance in terms of the backscattered current
$I_{\rm BS} = I_0-\langle I \rangle = (e^{2}/2\pi) V$, where $I_0
= (e^{2}/2\pi) U$. As can easily be shown, the lowest order
contribution to $I_{\rm BS}$ is of order $\lambda^{2}$. Without
loss of generality we assume $U>0$ and find, using the requirement
(\ref{relation}), that the backscattered current takes the form
\begin{equation}
I_{\rm BS} = \frac{e \lambda^{2}   g L}{4 v_F}
\int_{-\infty}^{\infty} d \tau \, e^{i u \tau} e^{4 \pi {\mathcal
C}(\xi_0,\xi_0,\tau)} \label{ibsresult}
\end{equation}
where $u=eU/  \omega_L$ is the ratio of the voltage to the
frequency $\omega_L=v_F/g L$ characterizing a ballistic traversal
of the wire of length $L$. Here, the correlation function
${\mathcal C}(\xi,\eta,\tau)=\langle \phi(\xi,\tau) \phi(\eta,0) -
\phi^2(\eta,0) \rangle_0$ in the absence of an impurity is
expressed in terms of the dimensionless variables $\xi=x/L$,
$\eta=y/L$ and $\tau=t v_F/gL$. At the dimensionless impurity site
$\xi_0$ one can show that
\begin{widetext}
\begin{eqnarray}
{\mathcal C}(\xi_0,\xi_0,\tau)= -\frac{g}{4 \pi} \left\{ \sum_{m
\in Z_{\rm even}} \gamma^{|m|} \ln{\left(\frac{ (\alpha+i
\tau)^2+m^2}{\alpha^2+m^2 }\right) } \, + \, \sum_{m \in Z_{\rm
odd}} \gamma^{|m|} \ln{\left(\frac{ (\alpha+i \tau)^2+(m-2
\xi_0)^2}{\alpha^2+(m-2 \xi_0)^2 }\right) } \right\}
\label{correlator}
\end{eqnarray}
\end{widetext}
where $\gamma = (1-g)/(1+g)$, and $\alpha=\omega_L/\omega_c$. Here
$\omega_c$ is the standard high-energy cutoff of the LL model with
$\lambda$, $eU$, $ \omega_L \ll \omega_c$.\\

\begin{figure}
\vspace{0.3cm}
\begin{center}
\epsfig{file=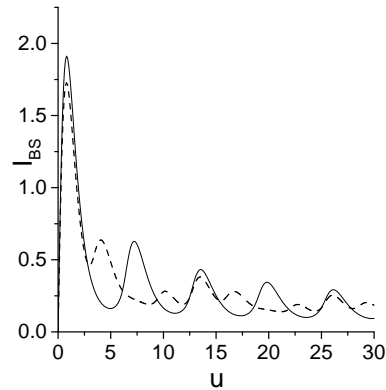,scale=0.3} \caption{\label{I_g025} The
backscattered current $I_{\rm BS}$ is depicted, in units of $e
(\lambda \omega_L^{\, g}/\omega_c^{\,g})^{2}/\omega_L {\cal
N}(\alpha;g)$, as a function of $u=eU/ \omega_L$. The interaction
strength is $g=0.25$. The solid line corresponds to $\xi_0 =0$ and
the dashed line to $\xi_0=0.25$. Decaying oscillations with a
constant period are clearly visible for the case $\xi_0=0$.}
\end{center}
\end{figure}

Using Eqs.~(\ref{ibsresult}) and (\ref{correlator}), the
backscattered current can now be evaluated numerically for
arbitrary values of the parameters $g$ (interaction strength), $u$
(voltage times length), and  $\xi_0$ (impurity position).
Fig.~\ref{I_g025} depicts the result as a function of $u$ for two
impurity positions (in the middle and at 1/4 of $L$), and for an
interaction value $g=0.25$ characteristic for single-walled
metallic carbon nanotubes \cite{egger97}. Apparently, the
backscattered current oscillates as a function of $u=eU/\omega_L$,
and the oscillations die out as $u$
increases.\\

Analytical expressions for the current can be obtained in special
situations. For a very long wire ($u \rightarrow \infty$) one can
perform an asymptotic expansion yielding
\begin{equation}
I_{\rm BS}\, = I^{\rm st}_{\rm BS}(U) \, [ 1\, + \, f^{\rm
osc}_{\rm BS}(u;\xi_0) \, + \, \ldots ] \label{IBSasym}
\end{equation}
where $I^{\rm st}_{\rm BS}$ is the leading order term while
$f^{\rm osc}_{\rm BS}$ describes the dominant oscillating
correction. The dots represent further subleading orders.
Explicitly one finds
\begin{equation}
I^{\rm st}_{\rm BS}(U)= \frac{\pi^2 \! {\cal N}(\alpha;g)}{
\Gamma(2g)}\, \frac{e^2 U}{2 \pi}\, \frac{ (\lambda
/\omega_c)^2}{(eU/\omega_c)^{2(1-g)}} , \label{ibsinf}
\end{equation}
which is independent of the length $L$ of the wire and the
impurity position, and exhibits a power-law behavior as a function
of the applied voltage $U$ in accordance with the result for the
homogeneous LL (see e.g.~\cite{kane92}). ${\cal N}(\alpha;g)$ is a
constant of order unity. In contrast, the corrections explicitly
depend on both $\xi_0$ and $L$ (through $u$), clearly showing that
oscillations are a finite-size effect. For $0 < |\xi_0| < 1/2$ the
asymptotic expansion yields
\begin{widetext}
\begin{equation}
f^{\rm osc}_{\rm BS}(u;\xi_0)= \frac{2 \Gamma(2g)}{\Gamma( g
\gamma)} \sum_{s=\pm} {\cal D}(g;s|\xi_0|)\,\frac{\cos{[ (1-2s
|\xi_0|) \, u \,- \, \pi g (1+s\gamma /2)]}}{u^{2g(1- \gamma/2)}}
\frac{(1+2s |\xi_0|)^{2 g \gamma} }{(1-2s |\xi_0|)^{g (2- \gamma)}
(16 |\xi_0|)^{g \gamma} } \; , \label{asy-exp-gen}
\end{equation}
\end{widetext}
where ${\cal D}(g;s\xi_0)$ is a numerical factor of order unity. A
similar asymptotic expansion has been performed in
Ref.~\cite{ponom97}, where the crossover from LL to FL
behavior was studied for a tunnel barrier.

The two frequencies appearing in the cosine terms of
Eq.~(\ref{asy-exp-gen}) are related to the distances of the
impurity from the two contacts (in units of the wire length $L$).
The phase shifts $u_s=\pi g (1 \pm \gamma/2)$ as well as the decay
factors of the oscillations directly depend on the
electron-electron interaction strength. Importantly, when the
interaction in the wire is switched off ($g\rightarrow 1$), i.e.,
when one deals with a homogeneous system, the oscillations vanish
since $\Gamma(g \gamma) \rightarrow \infty$ in this limit.
Clearly, $I_{\rm BS}$ also vanishes when there is no
back-scatterer, i.e. $\lambda \to 0$.

One can understand the origin of these current oscillations in
terms of interference effects of plasmon modes which are reflected
both by the impurity and, in an Andreev-type process
\cite{safi95}, by the wire-lead contacts. When varying either the
length $L$ of the wire, or the voltage (which appears as a phase
shift in the effective action (\ref{Seff})), one basically changes
the interference conditions, which results in the oscillatory
behavior of the current. The period of the oscillations is only
independent of $g$ if the backscattered current is considered as a
function of the dimensionless variable $u$; however, one has to
keep in mind that $u$ ($=eUgL/v_F$) has an intrinsic
$v_F/g$-dependence.
%%%%%%%%%%%%%%%%%%%%%%%%%%%%%%%%%%%%%%%
%%%%%%%%%%%%%%%%%%%%%%%%%%%%%%%%%%%%%%%
%%%%%      FIGURE    1          %%%%%%
%%%%%%%%%%%%%%%%%%%%%%%%%%%%%%%%%%%%%%%
%%%%%%%%%%%%%%%%%%%%%%%%%%%%%%%%%%%%%%%

We also point out that the expression (\ref{asy-exp-gen}) is
singular for $\xi_0 \rightarrow 0$; this is just due to a
mathematical complication when the distances of the impurity from
the two contacts become equal. Then pairs of poles in the
correlator merge. The full expression (\ref{ibsresult}) for the
current is however perfectly regular, and indeed one can still
calculate the asymptotic expansion for this particular impurity
position, obtaining
\[
f^{\rm osc}_{\rm BS}= \frac{\Gamma(2g) 2^{1-2 g \gamma}}
{\Gamma(2g \gamma)}
 \frac{\cos{[u-\pi g (1+\gamma)]}}{u^{2g(1- \gamma)}}
 \prod_{n=2}^{\infty} \left(\!\frac{n^2}{n^2-1}\! \right)^{2 g \gamma^{n}} .
\]
It should be emphasized that in the asymptotic expansion
(\ref{IBSasym}) the further subleading orders (represented by
dots) can be shown to become increasingly important the stronger
the interaction is, i.e., the lower the value of $g$. In practice,
retaining only $I^{\rm st}_{\rm BS}$ and $f^{\rm osc}_{\rm BS}$ is
a very good approximation for $g \gtrsim 0.5$ and $u \gtrsim 10$.
Nevertheless, also for $g < 0.5$ and/or lower values of $u$ (not
too close to $u=1$), Eq.~~(\ref{IBSasym}) gives a qualitatively
correct picture of the properties of the backscattered current as
we have checked numerically.

\begin{figure}
\vspace{0.3cm}
\begin{center}
\epsfig{file=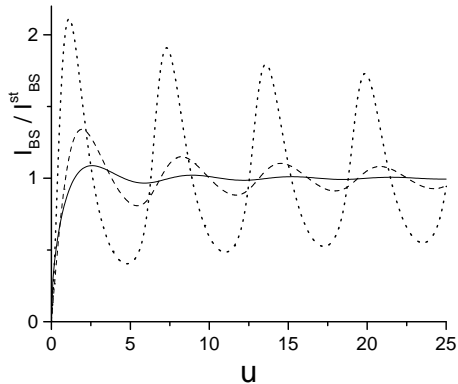,scale=0.35} \caption{\label{I_3} The
backscattered current $I_{\rm BS}$ is plotted, in units of $I^{\rm
st}_{\rm BS}$, as a function of $u=eU/ \omega_L$ for three
different values of the interaction strength: $g=0.25$ (dotted
line), $g=0.50$ (dashed line), and $g=0.75$ (solid line).
Electron-electron interaction affects the decay as well as the
phase shift of the oscillations.}
\end{center}
\end{figure}

We now discuss how the behavior of the backscattered current
changes with the interaction strength $g$. In Fig.~\ref{I_3} the
full numerical result of $I_{\rm BS}$, Eq.~(\ref{ibsresult}), is
plotted in units of $I^{\rm st}_{\rm BS}$, Eq.~(\ref{ibsinf}), for
three different values of $g$. For simplicity we have considered
the case of an impurity located in the middle. The figure
illustrates that the stronger the electron-electron interaction,
the more pronounced the oscillations are. This is due to the fact
that the Andreev-type reflections at the contacts are
interaction-dependent with a reflection coefficient
$\gamma=(1-g)/(1+g)$ \cite{safi95}. Thus, interference effects are
most pronounced in the limit $g \rightarrow 0$. Furthermore, we
see that the period of the oscillations is the same for any value
of $g$, namely $\Delta u =2\pi$ for $\xi_0=0$, but different
values of $g$ yield different phase shifts.

In the opposite limit of a short wire ($u \rightarrow 0$), the
integral (\ref{ibsresult}) can also be evaluated analytically, and
we find for the backscattered current
\begin{equation} \label{ibszero}
I_{\rm BS} = {\cal R}(\lambda,g) \frac{e^{2}}{2 \pi} U
\label{ibsusmall}
\end{equation}
with the cut-off and interaction dependent `reflection
coefficient'
\begin{equation} \label{ra}
{\cal R}(\lambda,g) = \pi^2{\cal N}_2(\alpha;g)
\frac{(\lambda/\omega_c)^2}{(\omega_L/\omega_c)^{2(1-g)}} \; ,
\end{equation}
where the constant ${\cal N}_2(\alpha;g)$ is of order unity. The
power law behavior of ${\cal R}(\lambda,g)$ as a function of the
system size $L$ agrees with the calculation of the linear
conductance from the Kubo formula \cite{ines}. Furthermore,
Eq.(\ref{ibszero}) bears a resemblance to the usual Landauer
formula. In the limit $g \rightarrow 1$ the reflection coefficient
approaches the expected value ${\cal
R}(\lambda,1)=\pi^2(\lambda/\omega_c)^2$.

%%%%%%%%%%%%%%%%%%%%%%%%%%%%%%%%%%%%%%%
%%%%%%%%%%%%%%%%%%%%%%%%%%%%%%%%%%%%%%%
%%%%%      FIGURE    2          %%%%%%
%%%%%%%%%%%%%%%%%%%%%%%%%%%%%%%%%%%%%%%
%%%%%%%%%%%%%%%%%%%%%%%%%%%%%%%%%%%%%%%

Finally we discuss the applicability of the perturbative approach.
To be consistent, the backscattered current $I_{\rm BS}$ has to be
small compared to the current $I_0$ in the absence of an impurity,
i.e.,
\begin{equation}
|I_{\rm BS} | \ll e^2 U/2\pi \; . \label{consistency}
\end{equation}
The analytical expressions obtained in the limits of both large
$u$, Eqs.~(\ref{IBSasym})-(\ref{ibsinf}), and small $u$,
Eqs.~(\ref{ibsusmall})-(\ref{ra}), suggest that the consistency
condition can be expressed in a simple and cut-off independent
way, if we introduce an effective impurity strength $\lambda^* =
\omega_c (\lambda /\omega_c)^{1/(1-g)}$. Indeed,
Eq.~(\ref{consistency}) then amounts to $\lambda^* \ll eU$ for
large $u$, and $\lambda^* \ll \omega_L$ for small $u$,
respectively. This is in accordance with general reasoning. It is
well known that the backscattering term (\ref{h_i}) is a relevant
perturbation to the LL fixed point, and therefore the RG-flow
would drive the system to two disconnected QWs, {\it unless} some
energy-scale cuts off the flow of $\lambda$. In our case the
applied voltage $U$ or the wire ballistic frequency $\omega_L$ can
play such a role. Therefore, as long as $\lambda^*$ is much
smaller than either $eU$ or $\omega_L$, the QW remains in the
neighborhood of the LL fixed point and our perturbative treatment
is justified. In the limit $g \rightarrow 1$, the effective
impurity strength $\lambda^*$ vanishes, and perturbation theory is
always applicable if $\lambda \ll \omega_c$.

In summary, we have investigated the backscattered current of an
interacting QW of finite size $L$ that is coupled to
non-interacting electron reservoirs. We have found that the
interplay between electron-electron interaction and finite size
gives rise to distinct features in the current-voltage
characteristics. In particular, we have found an oscillation of
the backscattered current as a function of $u=eU/\omega_L$. For
carbon nanotubes one can estimate a typical ballistic frequency of
the order of a few meV. From Figs.~\ref{I_g025} and \ref{I_3}, one
can see that the oscillations are visible for $eU \sim 10 \dots
50$ meV, a reasonable value, which is consistent with the cut-off
of about 1 eV. Hence, the predicted effects should be observable.

We thank H.~Bouchiat, R.~Egger, D.~C.~Glattli, and W. H{\"a}usler
for helpful discussions. Financial support by the CNRS, the DFG
and the EU and  are gratefully acknowledged.

%\bibliography{/home/wen/bib/wencross,/home/wen/bib/htc,/home/wen/bib/misc,/home/wen/bib/fqh,/home/wen/bib/part,/home/wen/bib/qcom,/home/wen/bib/publst}
%\end{multicols}
\end{document}